\documentclass[prl,twocolumn,showpacs,preprintnumbers,amsmath,amssymb,superscriptaddress]{revtex4}

\usepackage{graphicx}
\usepackage{dcolumn}
\usepackage{bm}

\begin{document}


\title{Photoluminescence of high-density exciton-polariton condensates}

\author{Natsuko Ishida}
\affiliation{CEMS, RIKEN, Wako-shi, Saitama 351-0198, Japan}

\author{Tim Byrnes}
\affiliation{National Institute of Informatics, Chiyoda-ku, Tokyo 101-8430, Japan}

\author{Tomoyuki Horikiri}
\affiliation{Department of Physics, Faculty of Engineering, Yokohama National University,  Yokohama, Kanagawa, 240-8501, Japan}
\affiliation{National Institute of Informatics, Chiyoda-ku, Tokyo 101-8430, Japan}

\author{Franco Nori}
\affiliation{CEMS, RIKEN, Wako-shi, Saitama 351-0198, Japan}
\affiliation{Physics Department, University of Michigan, Ann Arbor, MI 48104-4313, USA}

\author{Yoshihisa Yamamoto}
\affiliation{E. L. Ginzton Laboratory, Stanford University, Stanford, CA 94305}

\date{\today}

\begin{abstract}
We examine the photoluminescence of highly-excited exciton-polariton condensates in semiconductor microcavities. 
Under strong pumping, exciton-polariton condensates have been observed to undergo a lasing transition where the strong coupling
between the excitons and photons is lost.  
We discuss an alternative high-density scenario, where the strong coupling is maintained. 
We find that the photoluminescence smoothly transitions between 
the lower polariton energy to the cavity photon energy.  
An intuitive understanding of the change in spectral characteristics is given, as 
well as differences to the photoluminescence characteristics
of the lasing case. 
\end{abstract}

\maketitle

Exciton-polaritons are quasi-particle excitations in semiconductor microcavities consisting of a superposition
of a cavity photon and a quantum-well exciton.  These particles have been observed to undergo Bose-Einstein 
condensation (BEC) thanks to their very light effective mass, which they inherit from their photonic component \cite{Deng2002,Deng2006,Kasprzak2006,Lai2007,Deng2003,Balili2007,Ballarini2009,Cao1997,Jiang1998,Savvidis2001}. 
The achievement of exciton-polariton condensation has allowed investigating 
fundamental quantum states of matter, such as superfluidity~\cite{Amo2009} and quantized vortex formation in a semiconductor chip~\cite{Lagoudakis2008,Roumpos2011}. 
Currently, the low-density regime has been primarily investigated, 
where the average interparticle distance is much larger than the Bohr radius. 
In this regime, the polaritons are well approximated as bosonic particles, 
such that at sufficiently low temperatures BEC may occur. 
As the density is increased further beyond threshold, there has been experimental 
evidence that condensation crosses over into a lasing regime \cite{Deng2003,bajoni07,ballarini07,kammann12}.  This has been 
primarily interpreted as a result of exciton dissociation, which results in the loss
of strong coupling.  After strong coupling is lost, owing to the fact that the structure of the system is identical to 
a vertical cavity surface emitting laser (VCSEL), any coherence in the system is more
appropriately described as originating from photon lasing, rather than polariton condensation.  

However, the loss of strong coupling at high density is not the only possible scenario.  
We can also think of a high-density regime where temperatures are sufficiently low such that strong coupling
is maintained.  Such a strong-coupling high-density scenario has been the subject of many theoretical 
investigations~\cite{Keeling2005,Byrnes2010,Kamide2010,Eastham2001,Keeling2004,Szymanska2006,Szymanska2007,Yamaguchi2013}.
This regime is conceptually more complicated than the low density due to the underlying fermionic physics
of the excitonic constituents. The bosonic quasi-particle picture can no longer be applied, as phase-space-filling due to 
the Pauli exclusion principle gives a 
maximum density of excitons, while no such limit exists for photons.  This has led to discussions of 
whether the exciton-polariton BECs would undergo a crossover to photon-lasing based electron-hole plasma, 
or an electron-hole BCS-like phase. Another open question is how such phases would be probed experimentally, 
and what experimental signatures would distinguish the various high density scenarios.

\begin{figure}
\begin{center}
\includegraphics[width=8cm,clip]{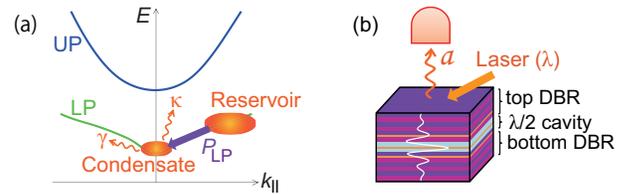}
\caption{(color online). (a) The model considered in this paper.  Lower polaritons (LP) are pumped into the condensate from the reservoir modes ($P_{{\rm LP}}$). 
They decay from the condensate with either the exciton decay rate $\gamma$ or the cavity photon decay rate $\kappa$.
(b) PL measurement scheme: A laser injected into the semiconductor sample in which the quantum wells are embedded. 
Photonic PL is defined as the photon leakage from the sample surface. DBR stands for distributed Bragg reflector.}
\label{fig:fig1}
\end{center}
\end{figure}

In this paper we investigate the photoluminescence (PL) of high-density exciton-polaritons.  The PL
is the most direct way of probing exciton-polaritons, and is desirable from an experimental point of 
view to know what differences in spectra may exist between the two scenarios, if any. We model the system as an 
open-dissipative system of hard-core (excitons) and soft-core (cavity photons) bosons with gain, loss, and an effective dephasing term
\begin{align}
\frac{d}{dt}\rho
= & 
-\frac{i}{\hbar}
\left[ \rho ,\mathcal{H} \right]
+
\frac{\kappa}{2}
\mathcal{L}(a,\rho)
+ \frac{P_{{\rm LP}}}{2}
\mathcal{L}(p^{\dagger},\rho)+\frac{\Gamma}{2} \sum_{i} \mathcal{L}(\sigma_i^z,\rho)
\label{eq:master}
\end{align}
where the Hamiltonian is 
\begin{align}
\mathcal{H/\hbar}
 = & \omega_0 {n}_{\rm ex} + \omega {n}_{\rm ph}
+ \frac{g}{\sqrt{M}} \sum_{i=1}^{M}   
\left( a^{\dagger} e_i +  a e_i^\dagger \right) \nonumber \\
& + \frac{U}{2} {n}_{\rm ex} ({n}_{\rm ex}-1)
\label{eq:hamiltonian_multi}
\end{align}
and $e_i^\dagger $ is the creation operator for an exciton at site $ i $ obeying 
bosonic commutation relations but having a maximum occupation of one $ (e_i^\dagger)^2 = 0 $, 
$a$ and $a^{\dagger}$ are annihilation and creation operators for the cavity photon, 
$ {n}_{\rm ph} = a^\dagger a $, $ {n}_{\rm ex} = \sum_{i=1}^M e_i^\dagger e_i $, and $ \sigma_i^z = 2 e^\dagger_i e_i - 1 $. The first two terms in 
 (\ref{eq:hamiltonian_multi}) are energy terms for excitons and photons, the third is the exciton-photon coupling, and $ U $ is a self-interaction energy between two excitons. 
Also, $ g $ is half the splitting between the lower polariton (LP) and upper polariton (UP) at low density, and 
$M$ is the total number of excitonic sites in the sample. 
The decay, gain, and dephasing terms are assumed to be of Lindblad form
$\mathcal{L}(O,\rho)=2 O \rho O^\dagger - O^\dagger O \rho - \rho O^\dagger O $.  
The decay rate for the cavity photon is $\kappa$ and the dephasing rate is $ \Gamma $. 
The gain term proportional to $ P_{{\rm LP}} $ pumps LPs according to 
$ p^{\dagger} \equiv \frac{1}{\sqrt{2}} \left ( \frac{1}{\sqrt{M}} \sum_i e^\dagger_i - a^{\dagger} \right ) $,
where we have assumed zero detuning $ \omega_0 = \omega $.  

The basis of such a model is as follows (see Fig.~\ref{fig:fig1}). The pump laser initially excites polaritons at 
high energy and momenta, which cool via phonon emission along the LP dispersion \cite{Deng2006}.  
As the polaritons cool to the vicinity of $ k = 0 $, a bottleneck develops where a large population of polaritons accumulate due to inefficient phonon cooling. 
Due to polariton-polariton interactions, the polaritons are now able to directly scatter into the condensate, with rate $ P_{{\rm LP}} $ \cite{schwendimann06,schwendimann08}. Due to exciton-photon coupling, the quasi-particle excitations at $ k=0 $ are approximated
as polaritons (rather than photons or excitons), hence the pumping is with respect to $ p $. To account for the effects of phase-space filling, the excitons have a hard-core nature with a maximum occupancy 
of $ M $, which is dependent on the sample size. Related models in the context of polariton condensation were considered in Refs.~\cite{Keeling2005,Eastham2001,Keeling2004}. 
In the context of quantum dots, similar models were considered in works such as Refs.~\cite{mu1992,Laussy11,Poddubny10,Moelbjerg13,Poddubny13}. However, these do not consider the above cooling mechanism leading to pumping of the condensate.  In these works generally the pumping is therefore with respect to excitons or photons, rather than polaritons, as we consider here.

\begin{figure}
\includegraphics[width=8.8cm]{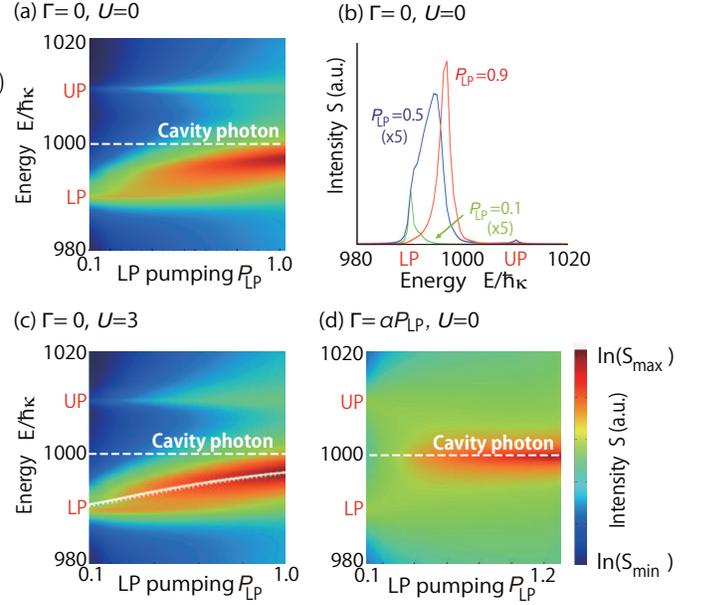}
\caption{(Color online) Photoluminescence (PL) of exciton-polariton condensates.  Parameters are chosen with
(a) (b) no dephasing $\Gamma=0$ and no interaction $ U = 0 $, (c) no dephasing  $\Gamma=0$ and with interactions $ U/\kappa = 3 $, and (d) with dephasing $\Gamma=\alpha P_{{\rm LP}}$, $\alpha=10$, $ U /\kappa= 0 $. Common parameters are $M=3$, $g/\kappa=10 $, and the cavity photon energy $\omega/\kappa=1000$ is labeled with the dashed lines. In (c), the mean PL energy for $U/\kappa = 3$ (solid line) and $ U /\kappa= 0 $ (dotted line) are shown for comparison. }
\label{fig:fig2}
\end{figure}

\begin{figure}
\includegraphics[width=8.5cm,clip]{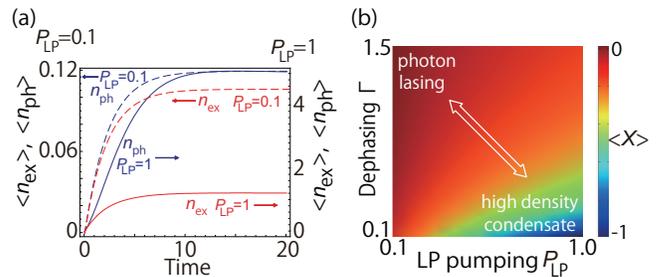}
\caption{(Color online) (a) The mean number of cavity photons $\langle {n}_{\rm ph} \rangle$ and excitons $\langle {n}_{\rm ex} \rangle$ for $\Gamma$=0.  (b) The exciton-photon coherence parameter $ \langle X \rangle = 
\frac{1}{\sqrt{M}} \langle a^\dagger \sum_i e_i + a \sum_i e_i^\dagger \rangle $. The parameters used are $g/\kappa$=10, $\omega/\kappa$=1000, and $M$=3. }
\label{fig:fig3}
\end{figure}

Figure \ref{fig:fig2}(a) shows the dependence of the PL on the LP pumping rate (see Supplemental Material for calculational methods).  
At low density only the usual LP is visible due to the direct pumping of the lower energy branch. 
As the density is increased, the overall behavior is that the peak PL shifts from the LP energy toward the cavity photon energy, consistent with previous calculations presented in Refs. 
\cite{Eastham2001,Byrnes2010}. The linewidth of the PL spectra is typically of the order of the photon decay rate $ \kappa $ at low densities, but increases significantly as the pump power is raised [Fig. \ref{fig:fig2}(b)].  
As the high-density regime is reached, the linewidth narrows again to the photon decay rate $ \kappa $, a behavior consistent with current experimental observations \cite{horikiri131,horikiri132} (See Supplemental Material showing 
additional experimental data). The mean number of cavity photons $\langle {n}_{\rm ph} \rangle$ and excitons $\langle {n}_{\rm ex} \rangle$ is shown in Fig. \ref{fig:fig3}(a).  
At low pumping powers and at steady state, we see that the mean numbers of photons and excitons are of the same order, as would be expected from the low-density 
polariton picture.
In the limit of zero density, there are exactly the same number of excitons and photons at zero detuning. 
At high densities, mean field theory predicts that the wavefunction approaches \cite{Byrnes2010}
\begin{align}
\exp \left[ \lambda a^\dagger-\lambda^2/2 \right] \prod_i \frac{1}{\sqrt{2}} ( 1- e^\dagger_i) |0 \rangle ,
\label{hidens}
\end{align}
where $ \lambda $ is the coherent amplitude of the light. 
Our numerics show that for high pumping rates the exciton number per site $\langle {n}_{\rm ex} \rangle /M $ approaches $0.5$, in agreement to this. 
In contrast, we see that the cavity photon number continues to increase with pumping strength. 
This is due to the soft-core nature of photons, which unlike the excitons, do not have to obey the Pauli exclusion principle. 
As the density increases further, the large photonic population starts to dominate the dynamics of the system. 
In Ref.~\cite{Byrnes2010}, this dominant photon population caused an effective binding of electrons and holes with a reduced Bohr radius. 

The shift of the PL spectrum from LP to cavity photon energy can be understood as follows. 
As the density approaches to and exceeds the Mott density ($n_{\rm Mott}=1/(\pi a_B^{2})$, where $a_B$ is the exciton Bohr radius), the photon population 
increases beyond the exciton population due to phase-space filling. 
Using the high-density mean-field wavefunction (\ref{hidens}), 
the energy of the high-density states can be evaluated to be 
$ E({n}_{\rm ph}) = \omega_0 M/2 + {n}_{\rm ph} \omega -g \sqrt{M {n}_{\rm ph}} + UM (M - 1)/8 $, 
where we have used the fact that $ \lambda^2 = {n}_{\rm ph}  $. 
As the PL emission corresponds to a loss of a single photon, let us consider the removal of a single photon from the system. 
The transition energy is then 
$\Delta E=E({n}_{\rm ph})-E({n}_{\rm ph}-1)=\omega-\frac{g}{2} \sqrt{\frac{M}{{n}_{\rm ph}}} \rightarrow \omega$, when ${n}_{\rm ph}$ is large. 
Thus as the system evolves towards high density, the peak PL approaches the cavity photon energy $\omega$. 
This general behavior holds even in the presence of polariton-polariton interactions [Fig. \ref{fig:fig2}(c)]. While there is a blue-shift to the spectrum
as the density increases, we still see the same general behavior where the peak PL energy shifts towards the cavity photon energy. Eventually the PL converges to the
cavity photon energy as the particles in the system are primarily photon-like [Fig. \ref{fig:fig2}(b)], 
which do not possess an interaction.

We note that despite the similarity of the model to resonance fluorescence, 
there is no characteristic of a Mollow's triplet spectrum.  The largest
factor which explains the lack of side peaks is the way in which the PL is being measured.  The PL
spectrum for exciton-polaritons is measured by accessing the light which escapes out of the microcavity. 
In general the photon decay rate is much faster than the exciton spontaneous decay.  
This means that the appropriate two-time correlation function to be calculated in the PL is between the cavity photons, and 
not the excitons. 
In order to reproduce the Mollow's triplet spectrum, it is necessary to evaluate the 
two-time correlation between the matter (or exciton in this case) operators, not the photon operators. 
As discussed in Refs. \cite{Byrnes2010,ishida13}, this gives a completely different set of transitions, 
which gives rise to side peaks, but are absent here.  
A secondary difference between the two scenarios is that resonance fluorescence starts in the weak-coupling regime, 
but in this case the excitons strongly couple to the photons even at low densities. 
Thus the transition from the LP energy to the cavity photon energy 
as seen in Fig.\ref{fig:fig2}(a) is absent in the Mollow's triplet spectrum, where the central
peak is pinned to the cavity photon energy. 
The evolution of the spectrum from the LP energy to the cavity photon energy is similar to that known from highly-excited quantum dots in the strong-coupling regime \cite{ishida13,Byrnes2010}. 
In comparison to the quantum dot case, the PL spectrum evolves from the LP energy to the cavity photon energy 
more smoothly, which is due to the increased number of possible optical transitions. This confirms the 
original prediction of Ref.~\cite{Byrnes2010} which was based on a mean-field calculation.

We now compare the PL to the lasing scenario described in the introduction. In this case, the strong laser pumping
adds a dephasing effect to diminish the strong coupling between excitons and photons, modeled by the term 
proportional to $ \Gamma $ in (\ref{eq:master}).  This may occur for example due to the presence of a large population of electrons and holes that are excited by the non-resonant laser.  
In a realistic experiment, it is likely that the amount of dephasing $\Gamma$ depends on the lower polariton pumping $P_{{\rm LP}}$. 
Increasing the pumping contributes to effects such as heating of the semiconductor sample which results in increasing $\Gamma$. 
To model this, we use a phenomenological relation $\Gamma=\alpha P_{{\rm LP}}$. 
We see in Fig. \ref{fig:fig2}(d) that this causes a discontinuous jump between LP and cavity photon energy, due to the increased dephasing. 
Thus, in this regime, excitons and photons co-exist with no superposition between the two. 
As the energy of a photon is then not modified from its original cavity photon energy, the PL emerges at this energy when the dephasing is large. 
This is in contrast to the smooth evolution of the peak PL without dephasing in Fig.\ref{fig:fig2}(a). 
In a photon-lasing scenario, one may wonder about the validity of the polariton pump model that we use in (\ref{eq:master}).  While it is more conventional to pump in either the exciton or photon basis in this regime, due to the large dephasing $ \Gamma $, polaritons are dephased immediately into half photons and excitons.  
Thus, regardless of the pumping scheme, photon lasing is achieved, and at equilibrium no exciton-photon superposition is present.

The general behavior of the system may be summarized by drawing a phase diagram as shown in Fig. \ref{fig:fig3}(b).  Here we plot the exciton-photon coherence parameter, defined as the expectation value at steady state of $ X= \frac{1}{\sqrt{M}} (a^\dagger \sum_i e_i + a \sum_i e_i^\dagger ) $.  
For a photon laser, we expect that this parameter is zero as photons and excitons are not present as a superposition in this regime. 
For a high-density polariton condensate, strong coupling can persist to high densities, which allows for this expectation value to take a non-zero value. 
We see that for large dephasing this parameter is zero, consistent with photon lasing, while large values are taken when the dephasing is small and strong pumping is present. This points to the presence of generally two possible phases, of photon lasing and a high-density polariton condensate, with a crossover connecting the two \cite{Yamaguchi2013}.  Strong pumping is seen to compensate somewhat for a large dephasing, which can be attributed to the large photon number reinforcing the coherence-generating term in (\ref{eq:hamiltonian_multi}) due to bosonic amplification.

The above cases have been restricted to the steady-state regime where a constant pump $ P_{{\rm LP}} $ is used.  
In a realistic experimental situation probing the high-density regime, a pulsed excitation is typically used in order to excite
the condensate due to limitations in the laser power \cite{horikiri131,horikiri132}.  
It is therefore a relevant question whether the inherent
transient dynamics of the pumping gives any qualitative differences to the PL spectra.  
In order to simulate the pulsed excitation, the LP pump profile is assumed to take the form
\begin{equation}
P(t) = \left\{ \begin{array}{ll}
P_{{\rm LP}}^{\rm max} \exp\left[ -\gamma t\right] & (t \geq 0) \\
0 & (t < 0)
\end{array} \right. .
\label{eq:pulse}
\end{equation}
Here $ \gamma $ is the decay rate of the excitons, and the zero of time is taken to be the moment the pulsed excitation is 
commenced. Typically, the duration of the pulsed excitation is extremely short ($\sim$ps), 
hence one may wonder why the relatively long timescale of $ 1/\gamma $ associated with the exciton lifetime is used in the exponential decay. 
After the initial excitation, excitons cool relatively slowly and accumulate in the reservoir, some of which may contribute to the condensate, and others decaying within the exciton lifetime.  Therefore, in terms of the pumping of the condensate in Fig. \ref{fig:fig1}, the reservoir exists for a time $ \sim 1/\gamma $ after which the reservoir depletes due to the finite lifetime of the excitons.  

\begin{figure}
\includegraphics[width=8.5cm]{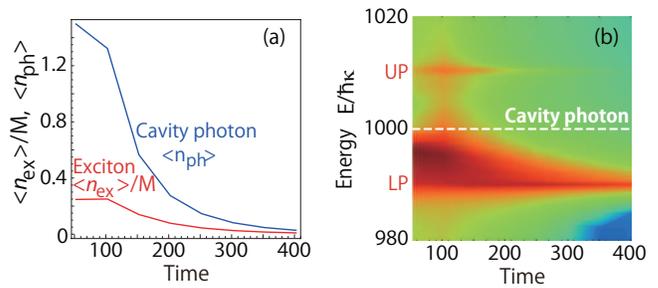}
\caption{(Color online). Time evolution of (a) the mean number of cavity photons $\langle {n}_{\rm ph} \rangle$ and excitons $\langle {n}_{\rm ex} \rangle /M $
and (b) the photoluminescence pumped by a time-dependent $P_{{\rm LP}}$. Dashed line indicates the cavity photon energy. 
The parameters used are $g/\kappa $=10, $\omega/\kappa $=1000, $\Gamma$=0, $\gamma/\kappa=0.01$, $P_{{\rm LP}}^{\rm max}$=0.88, $M$=3.
}
\label{fig:fig4}
\end{figure}

Figure \ref{fig:fig4} shows the PL for the pulsed-pump model. 
We see that the PL has a strong peak close to the cavity photon energy and decays towards the LP energy. 
The behavior is consistent with the results for constant pumping, 
but with a time dependence to the density.
The timescale of the transition is of the order of the exciton lifetime, which is a direct consequence of the pumping profile 
used.  The dynamics occurs on this timescale as the photon lifetime is much shorter than the exciton lifetime, which means that
the condensate quickly responds to changes in the pumping strength.  This explains the similar features seen in the constant-pumping
case, where a transition from the cavity photon to LP energy is seen.

In summary, we have investigated the photoluminescence properties of 
highly-excited exciton-polariton condensates. By assuming a model of LP pumping, 
we have shown that the PL shifts from the LP to cavity photon energy as it reaches the Mott density 
where the proportion of the cavity photons becomes dominant. This occurs without the 
loss of strong coupling between excitons and photons, in contrast to past interpretations
where such a transition was assumed to be a lasing transition.  Introduction of a
dephasing term between excitons and photons simulating the lasing case has a similar effect of 
pushing the PL towards the cavity photon energy. 
However, there is a distinct difference from the zero dephasing case 
that the PL jumps from the LP energy to the cavity photon energy at a certain point 
while the PL shifts smoothly for the zero dephasing case. Therefore, there are at least three mechanisms that can give rise to a blue shift in the spectrum: (i) polariton-polariton interactions; (ii) tendency of the PL to shift towards the cavity photon energy with high density; and (iii) dephasing reducing the Rabi splitting.  
In realistic systems, it is likely that a combination of all three effects plays a role to both the linewidth and the PL energy.  

The exciton-photon coherence parameter was
found to be a suitable observable distinguishing between a photon lasing and high density polariton condensate, 
where a crossover exists between the two regimes, depending upon the pump rate and the amount of dephasing. A time-dependent pumping profile was found to have similar qualitative results 
in probing the low to high density regime, by taking advantage of the relatively slow
decay of the excitons.  Due to the large difference in timescales of the exciton and photon lifetimes,
the system adapts rapidly to the changing densities due to the reservoir population.  While
we have based our calculations on a model with no underlying fermionic structure to the excitons, our calculations
using a BCS wavefunction have revealed qualitatively similar results, although for this case
a rigorous calculation of the PL is more difficult. We thus believe that many of the conclusions
would hold for either model, and the qualitative behavior is common either case.

We thank Makoto Yamaguchi, Yasutomo Ota, Yukihiro Ota, Kai Yan and Mike Fraser for discussions. 
This work is supported by the FIRST program of JSPS, Navy/SPAWAR Grant N66001-09-1-2024, Project for Developing Innovation Systems of MEXT, NICT, Transdisciplinary Research Integration Center, Okawa Foundation, Inamori Foundation, NTT Basic Laboratories, and JSPS KAKENHI grant Numbers 24740277 and 26790061. 
This work is also partially supported by the RIKEN iTHES Project, MURI Center for Dynamic Magneto-Optics, Grant-in-Aid for Scientific Research (S), and MEXT Kakenhi on Quantum Cybernetics.


\end{document}